\documentstyle[12pt]{article}
\parskip=\medskipamount

\newcommand {\cD}{{\cal D}}

\newcommand {\cL}{{\cal L}}

\newcommand{\bF}{{\bf F}}

\newcommand{\bH}{{\bf H}}

\def\a{\alpha}
\def\b{\beta}

\def\d{\delta}

\def\f{\phi}

\def\k{\kappa}
\def\l{\lambda}

\def\q{\theta}

\def\s{\sigma}

\def\z{\zeta}

\def\F{\Phi}
\def\J{\Psi}

\def\S{\Sigma}

\newcommand{\ad}{{\dot{\alpha}}}                           
\newcommand{\bd}{{\dot{\beta}}}                            
\newcommand{\ve}{\varepsilon}                            

\newcommand{\pa}{\partial}                           
\newcommand{\hf}{\frac12}

\newcommand{\be}{\begin{equation}}
\newcommand{\ee}{\end{equation}}
\newcommand{\bea}{\begin{eqnarray}}
\newcommand{\eea}{\end{eqnarray}}
\newcommand{\non}{\nonumber}

\begin{document}
\begin{titlepage}

\begin{flushright}
ITP-UH-24/97\\
hep-th/9709088 \\
\end{flushright}

\begin{center}
\large{{\bf SELF-INTERACTING VECTOR-TENSOR MULTIPLET} } \\
\vspace{1.0cm}

\large{Norbert Dragon and Sergei M. Kuzenko\footnote{ Alexander von
Humboldt
Research Fellow. On leave from Department of Quantum Field Theory,
Tomsk State University, Tomsk 934050, Russia.}
} \\
\vspace{5mm}

\footnotesize{{\it Institut f\"ur Theoretische Physik, Universit\"at
Hannover\\
Appelstra{\ss}e 2, 30167 Hannover, Germany} \\
dragon@itp.uni-hannover.de, kuzenko@itp.uni-hannover.de
 }
\end{center}
\vspace{1.5cm}

\begin{abstract}
We construct a consistent self-coupling for the vector-tensor multiplet
in $N=2$ harmonic central charge superspace. 
\end{abstract}
\vspace{15mm}


\vfill
\null
\end{titlepage}
\newpage
\setcounter{footnote}{0}


\noindent
The vector-tensor multiplet \cite{ssw} is an alternative off-shell
realization of the $N=2$ massless vector multiplet \cite{gsw}, in 
which one of the two physical scalar fields is converted into 
an antisymmetric tensor field. As a consequence, the vector-tensor
(VT) multiplet possesses $N=2$ supersymmetry with an off-shell
central charge generating the equations of motion. 
The VT multiplet has recently received considerable interest, since it 
describes the dilaton-axion complex of heterotic $N=2$ 
four-dimensional string vacua \cite{dkll} (see also \cite{sieb}). 

In contrast to $N=2$ vector multiplets, the consistent interactions for 
the VT multiplet turn out to be much more restrictive and inevitably 
require the presence of Chern-Simons forms. 
The Chern-Simons couplings of the VT multiplet to background $N=2$
vector multiplets have been classified, in the component approach,
in seminal papers of Claus et al \cite{cdfkst,cdft}. 
In essence, there exist two inequivalent versions whose specific features 
are linear and nonlinear central charge transformations. 
Recently, couplings of the VT multiplet to $N=2$ supergravity have been 
described \cite{sugra}.  

To fully understand the off-shell structure of the VT multiplet and its
interactions, a proper superfield formulation seems to be indispensable.
By now, three constrained superfield approaches to describe
the VT multiplet have been developed. In the $N=2$ central charge
superspace, the VT multiplet can be realized as a constrained
super 1-form \cite{how} or, equivalently, 
as a constrained super 2-form \cite{ghh,bho}. Finally, the VT multiplet
can be described in harmonic superspace \cite{ct,dk}.
In our opinion, the latter approach is most advantageous,
since the $N=2$ harmonic superspace \cite{gikos} is a universal framework
for general $N=2$ super Yang-Mills theories and $N=2$ supergravity.
It is almost obvious that an unconstrained superfield formulation 
for the VT multiplet may exist in harmonic superspace only.

In the present paper we continue the research started in \cite{dk}
and construct a consistent self-interaction for the VT multiplet.
This self-coupling is unique, modulo superfield
redefinitions, and leads to nonlinear central charge transformations.
The resulting model is closely related to that describing the VT multiplet 
coupled to a background abelian $N=2$ vector multiplet via gauging the
central charge \cite{cdfkst}. 
For a constant background vector multiplet (only the scalar fields
have non-zero constant values and the other fields vanish)
the bosonic action obtained in \cite{cdfkst} coincides with that 
corresponding to the self-interacting VT multiplet. 
Therefore, our superfield constraints
can be gauge covariantized, maybe in a rather complicated way,
to result in the VT multiplet with gauged central charge.   
\vspace{5mm}

\noindent
A superfield strength describing the VT multiplet lives over the $N=2$
central charge superspace \cite{s} 
with coordinates 
$\{ x^m,\, z,\, \theta_i^\alpha,\,\bar \theta^i_{\dot{\alpha}}\}$,
$\overline{\theta_i^\alpha}=\bar \theta^{{\dot{\alpha}}\,i}$
(with $z$ the central charge real variable) and covariant 
derivatives $D_M = (\pa_m, \pa_z, D^i_\a, {\bar D}_i^\ad)$
forming the algebra
\bea
& \{ D^i_\a, D^j_\b\} = -2{\rm i}\, \ve _{\a \b}\, \ve^{ij} \,\pa_z \qquad
\{ {\bar D}_{\ad i}, {\bar D}_{\bd j} \} = 2 {\rm i}\, \ve_{\ad \bd}\,
\ve_{ij}\, \pa_z \non \\
& \{ D^i_\a,{\bar D}_{\ad j} \} = -2{\rm i} \,\d^i_j \,\pa_{\a \ad} \;.
\label{1}
\eea  
On technical grounds, however, it is convenient to work in 
the $N=2$ harmonic central charge superspace \cite{gikos} 
which extends the above 
superspace by the two-sphere $S^2 =SU(2)/U(1)$ 
parameterized by harmonics, i.e. group elements
\bea
&({u_i}^-\,,\,{u_i}^+) \in SU(2)\non\\
&u^+_i = \ve_{ij}u^{+j} \qquad \overline{u^{+i}} = u^-_i
\qquad u^{+i}u_i^- = 1 \;.
\label{2}
\eea
Now, the set of covariant derivatives involves in addition the harmonic
derivatives
\be
D^{\pm \pm} = u^{\pm i} {\pa} / {\pa u^{\mp i}} \qquad
D^0 = u^{+i} {\pa} / {\pa u^{+ i}} -  u^{-i} {\pa} / {\pa u^{-i}}
\label{3}
\ee
where $D^0$ is the operator of $U(1)$ charge \cite{gikos}.
If one changes  the basis of covariant derivatives (and similarly 
for $\q$'s and $\bar \q$'s)
$$
D^\pm_\alpha = D^i_\alpha u^\pm_i \qquad 
{\bar D}^\pm_{\dot\alpha}={\bar D}^i_{\dot\alpha} u^\pm_i 
$$
the relations (\ref{1}) turn into
\bea
& \{ D^+_\a, D^-_\b \} =2{\rm i}\, \ve_{\a \b} \pa_z \qquad
\{ {\bar D}^+_\ad, {\bar D}^-_\bd \} =2{\rm i}\, \ve_{\ad \bd} \pa_z
\non \\
& \{ D^+_\a, {\bar D}^-_\bd \} = - \{ D^-_\a, {\bar D}^+_\bd \}
= - 2{\rm i}\,\pa_{\a \bd} 
\label{4}
\eea
and the other anticommutators vanish. 

To construct supersymmetric invariants in harmonic central charge
superspace, one can make use of the following action rule \cite{s,dk}
\be
S = \int du \,d\z^{(-4)} \,
\left( (\q^+)^2  - ({\bar \q}^+)^2 \right) \cL^{++} 
\label{5}
\ee
where
\be
 d\z^{(-4)}  =  \frac{1}{16} d^4 x \, D^{-\a}D^-_\a
{\bar D}^-_\ad {\bar D}^{-\ad}
\label{6}
\ee
and $\cL^{++}$ is a superfield of $U(1)$-charge $+2$ 
subject to the constraints
\be
D^+_\a \cL^{++} = {\bar D}^+_\ad \cL^{++} = 0 \qquad
D^{++} \cL^{++} = 0\;. 
\ee
The super Lagrangian can be equivalently rewritten in the form
\be
\cL^{++} = \cL^{ij} (x,z,\q) u^+_i u^+_j
\ee
for some $u$-independent superfields $\cL^{(ij)}$. The action is real
if $\cL^{++}$ is imaginary 
with respect to the analyticity preserving conjugation \cite{gikos},
in particular $\overline{\cL^{ij}} = - \cL_{ij}$.

The free VT multiplet is described by a real $u$-independent superfield
$L$ constrained by \cite{ghh,dk,bho}
\be 
D^{+\a} D^+_\a L = D^+_\a {\bar D}^+_\ad L = 0 \;.
\label{7}
\ee
The constraints imply that we have two possible 
candidates for the super Lagrangian
\bea
\cL^{++}_{{\rm vt,free}} &=& \frac{1}{4} \,(D^{+\a} L D^+_\a L - 
{\bar D}^+_\ad L {\bar D}^{+\ad} L ) 
\label{lag} \\
\cL^{++}_{{\rm der,free}} &=& \frac{{\rm i}}{4} \,
(D^{+\a} L D^+_\a L + {\bar D}^+_\ad L {\bar D}^{+\ad} L )\;. 
\label{der}
\eea
Only the former, however, describes the true Lagrangian of the VT 
multiplet, while the latter leads to a total derivative in components,
it generalizes the Chern form  and contains topological information only.

A consistent deformation of the constraints (\ref{7}) reads
\bea
D^{+\a} D^+_\a L &=& \k \, {\bar D}^+_\ad L {\bar D}^{+\ad} L
+ 2\k \, D^{+\a} L D^+_\a L \non \\
D^+_\a {\bar D}^+_\ad L &=& \k \, D^+_\a L {\bar D}^+_\ad L 
\label{8}
\eea
with $\k$ a real coupling constant of inverse mass dimension.
The deformed analogs of (\ref{lag}) and (\ref{der}) look like
\bea
\cL^{++}_{{\rm vt}} &=& \frac{1}{4} \, {\rm e}^{-3\k \,L} \,
(D^{+\a} L D^+_\a L - {\bar D}^+_\ad L {\bar D}^{+\ad} L )
\label{9} \\
\cL^{++}_{{\rm der}} &=& \frac{{\rm i}}{4} \, {\rm e}^{-\k \,L}\, 
(D^{+\a} L D^+_\a L + {\bar D}^+_\ad L {\bar D}^{+\ad} L )\;.
\label{10}
\eea
In contrast to the free case, the central charge transformations
become highly nonlinear
\bea
\pa_z D^+_\a L &=& - \pa_{\a \bd} {\bar D}^{+ \bd} L \non \\
& & - \frac{{\rm i}}{2}\k \, {\bar D}^{- \bd}
\left( D^+_\a L {\bar D}^+_\bd L \right) 
- \frac{{\rm i}}{4}\k \, D^-_\a 
\left(D^{+\b} L D^+_\b L + 2{\bar D}^+_\bd L {\bar D}^{+\bd} L \right) 
\non \\
& & + \frac{3}{4} \,{\rm i}\k^2 \, D^{--} 
\left(D^+_\a L {\bar D}^+_\bd L {\bar D}^{+\bd} L \right)
\label{cc}
\eea
as a consequence of the constraints.

Let us comment on derivation of the constraints and their
uniqueness. One could start with 
a general Ansatz for the deformed constrains
\bea
D^{+\a} D^+_\a L &=& a \,{\bar D}^+_\ad L {\bar D}^{+\ad} L
+ b \,D^{+\a} L D^+_\a L \label{11} \\
D^+_\a {\bar D}^+_\ad L &=& c \,  D^+_\a L {\bar D}^+_\ad L
\label{12}
\eea
where $a$ and $b$ are complex constants and $c$ a real parameter.
The parameters can not be completely arbitrary, since we should
guarantee the identity $(D^+)^3 L = 0$ 
along with the requirement that the two possible expressions
for ${\bar D}^+_\ad (D^+)^2 L$ derived from (\ref{11}) and (\ref{12})
respectively must coincide. These requirements lead to the two non-trivial 
solutions (there is also one trivial solution $a=0$, $b=c \neq 0$ describing
the free VT multiplet modulo a superfield re-definition)
\bea
& a = c = 0 \qquad b \neq 0 \label{13} \\
& | a | = \hf b = c \neq 0 \;.
\label{14}
\eea
There exist additional consistency requirements which have, however, 
more tricky and deeper origin. The simplest way to see how they come from 
is to turn to the component fields of $L$. The independent components
fields can be chosen as
\bea  
& \F = L| \quad \qquad D = \pa_z L| \non \\
& \l_\a^i = D^i_\a L| \quad 
\qquad {\bar \l}_{\ad i} = {\bar D}_{\ad i} L| \non \\
& F_{\a \b} = \frac{1}{4} [D^+_{(\a} \, , D^-_{\b)} ] L| \qquad
\quad {\bar F}_{\ad \bd} = - \frac{1}{4} [{\bar D}^+_{(\ad} \, , 
{\bar D}^-_{(\bd)} ] L| \non \\
& H_{\a \ad} = {\bar H}_{\a \ad} = 
\frac{1}{4} [D^+_\a \, , {\bar D}^-_\ad ] L|
- \frac{1}{4} [D^-_\a \, , {\bar D}^+_\ad ] L| \;.
\label{15}
\eea
In the free case, the fields $H_m$ and $F_{[mn]}$ satisfy the constraints
\be
\pa^m H_m = 0 \qquad  \ve^{mnrs} \pa_{n} F_{rs} = 0
\label{16}
\ee
which are solved by
\be 
H^m = \hf \ve^{mnrs} \pa_n B_{rs} \qquad F_{mn} = 
\pa_{m} V_{n} - \pa_{n}V_{m}
\label{17}
\ee
with the gauge 1-form $V_m$ and 2-form $B_{mn}$ being unconstrained.
Therefore, any consistent deformation of the free constraints must
preserve the number of the off-shell degrees of freedom. This is a strong
requirement which the solution (\ref{13}) does not meet. As to the 
solution (\ref{14}), it turns out to be compatible with the 
consistency requirement under the additional restriction $a = c$.
It is worth pointing out that the $a$-structure in (\ref{11}) is 
responsible for the Chern-Simons coupling at the component level.
Without presence of this structure the deformed constraints are physically 
inconsistent. 

One could start with more general Ansatz for the deformed constraints,
than we have considered, namely
with $L$-dependent coefficients $a$, $b$, $c$ in eqs. (\ref{11}) and
(\ref{12}). But all the solutions for such an Ansatz prove to be
equivalent to those already encountered modulo superfield
re-definitions. Therefore, the constraints (\ref{8}) describing the 
self-interacting VT multiplet are unique.  

In the case of self-coupling (\ref{8}) the analogs of relations (\ref{16})
look like
\bea
\pa^m \left(H_m - \frac{3}{4} \k \, \S_m \right) &=&
\frac{1}{4} \k \, \left( {}^*F^{mn} - \frac{1}{4} \k \, {}^*\J^{mn} \right)
\left(F_{mn} - \frac{1}{4} \k \,\J_{mn} \right) \non \\
& & + 2 \k \, \pa^m \F \cdot \left(H_m - \frac{3}{4} \k \, \S_m \right)
\label{18} \\
\pa^m \left( {}^*F_{mn} - \frac{1}{4} \k \, {}^*\J_{mn} \right) 
&=&  \k \,\pa^m \F \cdot 
\left( {}^*F_{mn} - \frac{1}{4} \k \, {}^*\J_{mn} \right) 
\label{19}
\eea
where we have introduced
\be
\S_m = \l^i \s_m {\bar \l}_i \qquad 
\J_{mn} = - \l^i \s_{mn} \l_i + {\bar \l}_i \,\tilde{\s}_{mn} {\bar \l}^i 
\label{20}
\ee
with the $\s$-matrices defined as in \cite{bk}. From (\ref{18}) and (\ref{19})
we deduce
\bea
H_m &=& {\rm e}^{2\k \, \F} \bH_m + 
\frac{3}{4} \k \, \l^i \s_m {\bar \l}_i \non \\
F_{mn} &=& {\rm e}^{ \k \, \F} \bF_{mn} 
- \frac{1}{4} \k \,\left(\l^i \s_{mn} \l_i -
{\bar \l}_i \,\tilde{\s}_{mn} {\bar \l}^i \right)
\label{21}
\eea
where $\bH_m$ and $\bF_{mn}$ are constrained by
\be
\pa^m \bH_m = \frac{1}{4} \k \, {}^*\bF^{mn}\, \bF_{mn} \qquad
\ve^{mnrs} \pa_{n} \bF_{rs} = 0 \;.
\label{22}
\ee
These equations are readily solved in terms of unconstrained gauge
1-form $V_m$ and 2-form $B_{mn}$ 
\be
\bH^m = \hf \ve^{mnrs} \left( \pa_n B_{rs} +  \k \,
V_n \pa_r V_s \right) \qquad \bF_{mn} = \pa_m V_n - \pa_n V_m \;.
\label{23}
\ee
We see that the gauge fields interact via the Chern-Simons form.

Now, let us turn to the supersymmetric invariants generated by
the super Lagrangians (\ref{9}) and (\ref{10}). Similarly to the free
case, $\cL^{++}_{\rm der}$ turns out to produce a total derivative
in components. Eq. (\ref{9}) defines the Lagrangian of the self-interacting
VT multiplet. Rather lengthy calculations lead to the following
component Lagrangian
\bea
\cL_{\rm vt} &=& \hf \, {\rm e}^{-3\k \, \F} \left( \pa^m \F \pa_m \F
+ D^2 \right)  \non \\
& & - \frac{1}{4} \, {\rm e}^{-\k \, \F} \bF^{mn} \bF_{mn} 
- \hf \,{\rm e}^{\k \, \F} \bH^m \bH_m  \non \\
& & + \frac{1}{4} \k \, {\rm e}^{- 2 \k \, \F} \bF^{mn} \J_{mn}
- \hf \k \, {\rm e}^{-\k \, \F} \bH^m \S_m  \non \\
& & - {\rm e}^{-3\k \, \F} \left(  \frac{{\rm i}}{2}
\l^i \s^m \stackrel{\leftrightarrow}{\pa}_m {\bar \l}_i
- \frac{1}{32} \k^2 \, \J^{mn} \J_{mn}
+ \frac{3}{16} \k^2 \, \S^m \S_m \right)   
\label{24}
\eea
with $\S_m$ and $\J_{mn}$ given in (\ref{20}). If we re-define 
the scalar fields by 
\be
\f = {\rm e}^{-\k \, \F} \qquad \cD = {\rm e}^{-\k \, \F} D
\label{25}
\ee
the bosonic sector of $\cL_{\rm vt}$ turns into
\be
\cL_{\rm vt,bos} = \hf \, \f \left( \frac{1}{\k^2}  \pa^m \f \pa_m \f
+  \cD^2 \right)  
- \frac{1}{4} \, \f \, \bF^{mn} \bF_{mn}
- \hf \,\f^{-1} \, \bH^m \bH_m \;.
\label{26}
\ee
This expression coincides with the bosonic action of the VT multiplet
with gauged central charge \cite{cdfkst} for the case when the scalar
fields of the background $N=2$ vector multiplet have constant non-zero
values and the other fields vanish. Therefore, a superfield description
of the VT multiplet with gauged central charge can be obtained by
properly gauge covariantizing the constraints (\ref{8}) of the 
self-interacting VT multiplet. The naive covariantization of the
constraints (\ref{8}), which
consists in replacing the flat covariant derivatives by
gauge covariant ones (with respect to local central charge transformations),
is most likely inconsistent and some non-minimal terms depending
on the vector multiplet's superfield strengths should be added.
This problem is now under investigation.

Of course, the main reason why any time we should carry out a complicated
analysis on consistency of the constraints lies in the fact that we work
with constrained superfields. These problems would never have appeared
if we had an unconstrained superfield formulation for the VT multiplet.
Only such formulations makes the superfield approach much more powerful
than the component one. Similarly to the Fayet-Sohnius hypermultiplet, 
unconstrained
superfield formulation(s) for the VT multiplet could exist in harmonic
superspace only. Finding such a formulation seems to be 
one of the urgent problems of supersymmetric field theory.         
\vspace{5mm}

\noindent
After this work had been reported at the $31^{\rm st}$ International
Symposium Ahrenshoop on the Theory of Elementary Particles in Buckow 
we became aware that similar results were independently
obtained by E.A. Ivanov and E. Sokatchev \cite{is}.

\vspace{5mm}

\noindent
{\bf Acknowledgements.}
We are grateful to I.L. Buchbinder, B. de Wit, S.J. Gates,
E.A. Ivanov, B.A. Ovrut, E. Sokatchev and U. Theis
for fruitful discussions.
This work was supported by
the RFBR-DFG project No 96-02-00180,
the RFBR project No 96-02-16017 and by the Alexander
von Humboldt Foundation.


\end{document}